# Direct evidence of photo-ionization causing luminescence "off" state in CdTe Quantum dots

A.V. R. Murthy, Padmashri Patil, Shouvik Datta, Shivprasad Patil*

Division of Physics, Indian Institute of Science Education and Research,

Pune - 411008, Maharashtra, India

**\* Corresponding Author: Email: s.patil@iiserpune.ac.in**

## Abstract

We present a direct experimental evidence of photo-ionized CdTe quantum dot core having luminescence "off" state at band-edge photo-excitation. We use the autocorrelation of fluorescence fluctuations from dilute colloidal solution of CdTe quantum dots in a diffraction limited detection volume to determine their photo-ionization probability. The ionization probability is in the same order of magnitude ($\approx 10^{-6}$) as the CdSe and CdSe/ZnS core/shell quantum dots. Further, we measured the ionization/ neutralization rates ($\approx 10^{4}$ s$^{-1}$) after addition of β-mercaptanol and found that ionization rate is enhanced by two orders of magnitude after the addition. However, a comparable neutralization rate enhances the light emission quantum yield. The ionization/ neutralization at microsecond time scales puts an upper limit on achieving better quantum yield by thiol addition. The results also indicate that the dominant ionization mechanism at band-edge excitation is tunneling.

Optical properties of colloidal semiconductor quantum dots (QDs) are important both for harvesting solar energy and using them as high quantum yield biomarkers [1-4]. The central process that determines the use of QDs in these applications is their interaction with light. The proposed picture of light-QD interaction is that - a) an absorbed photon may cause ionization by transferring the electrons to surface trap states or the surrounding matrix with a very small probability $\approx 10^{-6}$ [5-6]; b) once charged, the QD does not emit light and the emission quantum yield is reduced [7]. The additional charge carriers in the core is thought to be responsible for this photo-luminescence quenching, whereas ionization itself is usually considered to be an Auger process involving two photons. [8-11]. The photo-ionization involving the surface states is experimentally established by measuring the photo-induced charge on individual QDs residing on a surface [6, 12]. However, the mechanism of photo-ionization and the role of ionized core in the loss of luminescence are often debated [12-18]. Recently, the charging model for loss of luminescence, for which there is no direct experimental evidence yet, is challenged by Zhao et al. [17]. Additionally, it is not yet clear if the charge transfer is thermally activated process, or a two photon process like Auger ionization or it is facilitated by tunneling through surface barrier [5-6, 13-16]. So far, various approaches such as Electrostatic Force Microscopy (EFM) [5, 6], the "on" and "off" time distribution of the luminescence from a single QD [13-16] and time-resolved photoluminescence measurements [19] are used to investigate photo-ionization in QDs. A photo-excited electron-hole pair recombines with either radiative or non-radiative pathways depending on the rates of these processes [11, 20]. A third possibility of charging also occurs, wherein either a hole or an electron is taken away from the core by surface traps or ligands [13-16]. The positively or negatively charged QD core is thought to suppress the radiative decay rates and lowers quantum yield significantly [21]. A single QD then exhibits blinking having a power-law distribution of "on" and "off" times and a reduced ensemble photoluminescence [13-16]. It was found that the addition of thiols such as β-mercaptanol (BME) to the solution of CdSe/ZnS core/shell QDs suppresses such blinking [22, 23].

In this letter, we propose a new method of measuring the fraction of ionized, non-luminescent QDs and also the ionization rates using Fluorescence Correlation Spectroscopy (FCS) for band edge photo-excitation. It is observed for the first time that photo-induced ionization of QDs in aqueous solution is possible even at band-edge excitation with a probability $\approx 10^{-6}$. Moreover, this number is in agreement with EFM measurements and luminescence "on" time distribution for QDs residing on a surface [5, 6]. The results are first and arguably direct evidence for the charging model of luminescence "off" state. The fraction of ionized, non-luminescent QDs varies linearly with photo excitation intensity. This also supports the view that the ionization at band-edge photo-excitation happens via a one photon process [6] and not a two-photon Auger-like event. Using FCS it was possible to measure the "on-off" transition rate in the presence of BME. The dependence of this rate on QD size and its insensitivity to temperature confirms often-suggested picture of ionization by tunneling through surface barrier. Further, it was found that the rate of ionization in the absence of BME is two orders smaller than in the presence of BME. We argue that the BME addition does not achieve suppression of blinking by restricting the charge transfer. Instead, it pushes the ionization-neutralization rate to micro-second times-scale which is not captured in milli-second binning times used in the past. However, a charge neutralization rate comparable to the ionization rate results in better quantum yield.

Fluorescence Correlation Spectroscopy (FCS) is conventionally used to measure the chemical kinetics of reactions involving "on" and "off" fluorescence states, diffusion of colloidal particles in various fluid mediums and the number of fluorophors in the detection volume [24-28]. In all these measurements, it is the autocorrelation of equilibrium fluctuations that gives information about the system under consideration. In the present set of experiments a very dilute aqueous solution of CdTe QDs ($10^{16}$ to $10^{17}$ particles/liter) capped with Mercapto Succinic Acid (MSA) [29] is excited with a 532 nm laser (Dream lasers, China). The detection volume is of the order of femto-liters produced by a diffraction limited focusing of the laser beam using a water immersion objective with high numerical aperture (60X, 1.2NA, Olympus, Japan). The emitted light is then focused onto a pinhole of size 25 µm. The excitation and emission are separated from each other using dichroic filter and an emission filter (Omega optics) [29]. The concentration of CdTe QDs is such that on average there are less than 10 QDs in the detection volume. The intensity of the laser is varied using a combination of neutral density filters. The autocorrelations are calculated using the fluorescence fluctuations recorded by an Avalanche Photo-detector (APD) (SPCM Module, Perkin Elmer). Unlike other attempts to measure hydrodynamic size of QDs using FCS [30, 31], we fit a standard diffusion model without considering an explicit term for the blinking, which is given by

$$G(\tau) = \frac{1}{N\left(1 + \frac{\tau}{\tau_D}\right)\left(1 + (r/l)^2 \frac{\tau}{\tau_D}\right)^{1/2}} \qquad (1)$$

Here N is the average number of emitting particles and $\tau_D$ is the average time spent by the particles in the detection volume also called residence time, $r/l$ determines the shape of the detection volume. It is well established that QD's exposure to the light results in photo-ionization with ionization probability ≈ $10^{-6}$ [6, 12]. If photo-ionization causes loss of luminescence, then the concerned QD stays dark for duration longer than the average time required to cross the detection volume in FCS experiment. The loss of fluorescence half way through its trajectory in the detection volume results in intensities correlating over shorter lag times. Fitting equation 1 in this situation is expected to give apparent average residence time which is less than the actual $\tau_D$. Figure 1(a) shows a representative autocorrelation curve fitted to equation 1 at 11 kW/cm$^2$. We also observe in Figure 1(b) that the measured $\tau_D$ does not vary much with light intensity below 10 kW/cm$^2$ and it is 220 µs. Above this critical intensity, $\tau_D$ starts to decrease and becomes even smaller than Rhodamine, a standard organic fluorophore (~60 µs). As shown in figure 1(b), $\tau_D$ for Rhodamine does not vary with intensity. The $\tau_D$ is related to hydrodynamic radius through $R_H = (2KT/3\pi\eta r^2)\tau_D$, where η is viscosity of the medium and K is Boltzman constant. The constant $\tau_d$ below 10kW/cm$^2$ translates into hydrodynamic radius of 3 nm. This is close to the actual average size of these QDs estimated using effective-mass approximation to their excitonic absorption spectra. Clearly, there is no intermittency in QD luminescence below this excitation intensity and autocorrelation of fluorescence fluctuations gives a correct hydrodynamic size.

Fitting equation 1 to the autocorrelation of fluorescence fluctuations can also give the average number of fluorophores N. We see in figure 2(a) that above 10 kW/cm$^2$ the average number of QDs gradually increases to a saturation value after addition of controlled amount of β-mercaptanol (BME). This is also accompanied by the rise in $\tau_D$ to a value close to 200 µs. It follows a sigmoidal growth with respect to BME concentration. It has been reported that the BME addition to the solution of colloidal QDs removes intermittency in the luminescence time-trace and the QD luminescence appears continuous. Figure 2 (a) also shows that the luminescence of all the dark QDs can be recovered with more than 70 µM concentration of BME in the solution. We measured the non-luminescent fraction by measuring the N with and without BME addition. This dark fraction is $\Delta N/N = (N - N_b)/N$. Here $N_b$ is the number of bright quantum dots before addition of BME and N is the number of QDs after addition of BME. Interestingly, as shown in figure 2 (b) this fraction increased linearly with excitation intensity. An electron-donating reducing agent such as BME helps in neutralizing and recovering this dark fraction indicates that ionization is responsible for the "off" state of the fluorescence. We refer to this photo-darkened fraction as photo-ionized fraction. We further estimate the photo-ionization probability by measuring the photo-induced non-luminescent fraction at different laser intensities. The measured absorption cross section ($\rho$) for CdTe quantum dots used in this experiment is $2 \times 10^{-16}$ cm$^2$. If I is the intensity in W/cm$^2$ then the number of photons absorbed in time t, is $I\rho t/h\nu$. Multiplying this number by photo-ionization probability k gives us photo-ionized fraction; $\Delta N/N = kI\rho t/h\nu$. We have taken exposure time t to be $\tau_d$ at intensity 5 kW/cm$^2$. We fit a straight line to the experimental data in figure 2 (b) and the probability k can be determined from the slope; $k = 9 \times 10^{-6}$. This implies that, on an average a QD needs to absorb $10^6$ photons before it photo-ionizes and enters into an "off" state.

Colloidal semiconductor quantum dots in the solution phase do not show photo-ionization with band edge excitation for intensities close to few mW/cm$^2$ [19]. However significant photo-ionization for immobilized dots is seen for intensities around 10 kW/cm$^2$ [12]. Our measurements also show that CdTe QD can enter a photo-ionized dark state in aqueous solution above this intensity with probability $9 \times 10^{-6}$. This is same order of magnitude compared to TOPO capped CdSe QDs immobilized on SiO$_2$ surface [6] and CdSe/ZnS core/shell QDs on quartz surface [5]. These experiments are performed under similar intensities. It is noteworthy that ionization probability is in the same order of magnitude for such a variety of QDs placed in widely different environments.

We also note that there is no considerable effect of BME addition on the number of QDs below 10 kW/cm$^2$. This particular threshold light intensity corresponds to $5 \times 10^6$ photons absorbed per QD per second. The QDs are under illumination for less than 220 µs of average residence time. Clearly, below 10kW/cm$^2$ the number of photons absorbed by a QD in its transit through the detection volume is less than $1.1 \times 10^3$. A QD rarely ionizes since the ionization probability is $9 \times 10^{-6}$. At the highest intensity ($4 \times 10^2$ kW/cm$^2$) used in these measurements the number of photons absorbed by the QD while in the detection volume is $4.5 \times 10^5$. The probability for photo-ionization is $9 \times 10^{-6}$. This means that roughly five ionization events per QD are possible while in the transit. Thus the threshold intensity or the range of intensity required for ionization is explained by ionization probability estimated using the ionized fraction measurement.

We further discuss the charge transfer dynamics in CdTe QDs by measuring on-off relaxation rates at equilibrium after BME addition. In equilibrium, if chemical reaction causes the fluorescence fluctuations along with diffusion, a reaction-diffusion model is fitted to the measured autocorrelation curves [24-26]. If we consider the reaction responsible for fluorescence fluctuations to be the exchange of electrons between the core and the surface states, a fit of reaction-diffusion model yields a reaction rate or "on-off" rate $k_r = 1/\tau_r$. We refer to this as charge transfer rate. The equation for fitting such reaction-diffusion model to the experimental autocorrelation curve is [26]

$$G(\tau) = \left(1 + \frac{F}{1-F} e^{-\tau/\tau_r}\right) \frac{1}{N\left(1 + \frac{\tau}{\tau_D}\right)\left(1 + \left(r/l\right)^2 \frac{\tau}{\tau_D}\right)^{1/2}} \qquad (2)$$

Here the net relaxation rate to equilibrium can be composed of two parts so that $k_r = k_i + k_n$, and $k_r = 1/\tau_r$. The $k_i$ and $k_n$ are ionization and charge neutralization rates representing a forward and backward transfer of electrons from core to the surface trap states. The dark and bright fractions at equilibrium are given by F and 1-F respectively. The fitting of equation (2) to the autocorrelation data at 25kW/cm$^2$ intensity in presence of BME yields a single exponent $\tau_r$ = 35 ±2.1 µs, F = 0.41 ±0.02, N= 6 ± 0.2 [29]. These values of fitting parameters indicate that, at equilibrium, the QD's core exchanges electrons with surface trap states at a single rate of ≈10$^4$ and about 40 percent particles are dark at any given time. It has been established that QDs in solution or on SiO$_2$ substrate have charge neutralization rates of the order of one per second [5, 6, 19]; However we find that in the presence of BME, both ionization and neutralization rates are 10$^4$ s$^{-1}$. As this blinking on microsecond time-scale is not captured in single QD luminescence experiments using millisecond binning times, the QD luminescence appears as continuous or non-blinking.

We recorded autocorrelation curves at different excitation intensities after BME addition. The fitting of equation 2 to this data, gives $k_r$ at different intensities [29]. Figure 3 shows dependence of $k_r$ on the excitation rate. The linear dependence on excitation rate again confirms that the observed photo-ionization by charge transfer is actually a one-photon process. The slope of this graph is 2.4× 10$^{-4}$. There is roughly one electron transfer per 10$^4$ absorbed photons, which means that ionization probability after BME addition is ~10$^{-4}$. However, the ionization probability deduced from the ionized fraction in figure 2(b) and also from EFM measurements by Li et al. is ~ 10$^{-6}$. This is an enhancement by two orders of magnitude. On the other hand, it is well known fact in the literature and from our own measurements that BME addition suppresses blinking and enhances photoluminescence. To explain this, we propose that addition of BME opens up faster channels for both ionization and charge neutralization. The relaxation times $\tau_r$ (~30 µs) to reach an equilibrium dark/bright ratio (~ 0.4) indicates that the QDs are blinking at the µs time-scales after addition of BME. Unlike tuning the barrier shape for non-blinking quantum dots [32], the BME addition actually does not appear to suppress any electron transfer. This is expected; since BME assisted charge neutralization at the QD surface actually reduce the surface barrier height of these QDs. It enhances both the ionization/neutralization rate in order to maintain the equilibrium charged fraction at any given time. Figure 4 shows the schematic of this process. Clearly, it is better to tune the barrier shape [32] in order to enhance the quantum yield rather than addition of thiol because the later does not actually suppress charge transfer at all time scales. A certain fraction of QDs always remains dark at equilibrium, putting an upper limit on enhancement of quantum yield.

The inset of figure 3 shows that $k_r$ does not depend on temperature in the modest range of 25 to 70 $^o$C. The insensitivity of $k_r$ to the sample temperature and its dependence on excitation rate suggest that tunneling is the dominant charge transfer mechanism. The QD diameter (well width for the photo-generated electron) affects the attempt frequency for tunneling the surface barrier and consequently the charge transfer rates. The attempt frequency for tunneling actually scales as inverse square of the QD diameter [33, 34]. If the charge transfer is mostly governed by tunneling then its rate should have an inverse square dependence on QD diameter [33]. Indeed, the $k_r$ measured for different sized QDs vary as inverse square of the QD diameter, supporting the tunneling theory of ionization [29].

In conclusion, we propose a new method to determine the photo-ionized fraction of QDs under band-edge excitation using FCS. The results establish that the photo-ionization causes loss of fluorescence. The measurements further suggest that addition of BME has the effect of enhancing the neutralization rates and this suppresses blinking on millisecond time-scales but QDs luminescence fluctuates at microsecond time scale. This new analysis of experiments performed using FCS to investigate the light-QD interaction can help resolve many issues in physics of light emission from nanostructures in general. The method will be useful in characterizing and further enhancing the emission quantum yields from various nanostructures.

The research is funded in part by Department of Science and Technology (DST) nano-science unit grant (SR/NM/NS-42/2009) and funding from Department of information technology (DIT) grant (No. 12(4)/ 2007-PDD) and from IISER-Pune.

**Figures and Figure captions:**

**Figure 1**

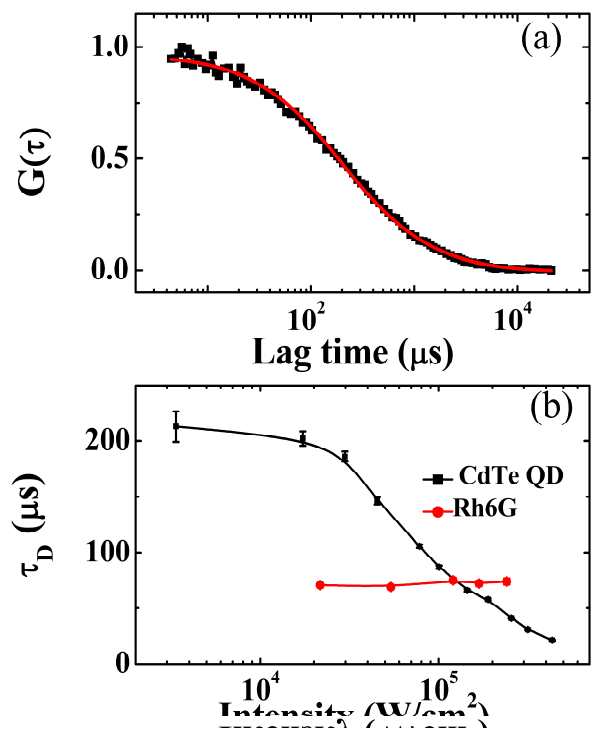

**Figure 1:** (a). A representative autocorrelation curve fitted to equation 1 to determine N and $\tau_D$. (b) The dependence of apparent average residence time $\tau_D$ on the excitation intensity. For larger intensities the apparent $\tau_D$ is even smaller than Rhodomine indicating a photo-assisted darkening as described in the text.

**Figure 2**

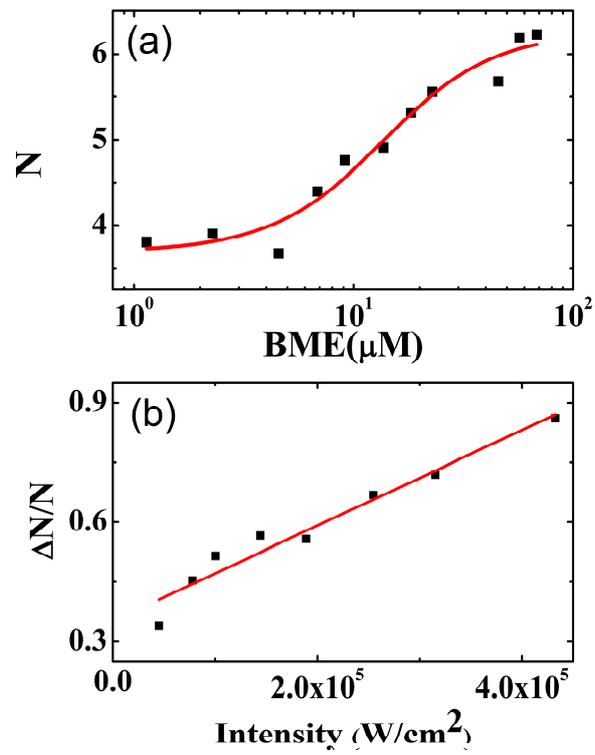

**Figure 2:** (a) The effect of gradual addition of BME to the solution on N. The continuous line is a fit to the sigmoid growth. (b) The charged fraction ΔN/N verses intensity. The ionization probability from the slope of the linear fit is 9.4 X $10^{-6}$.

**Figure 3.**

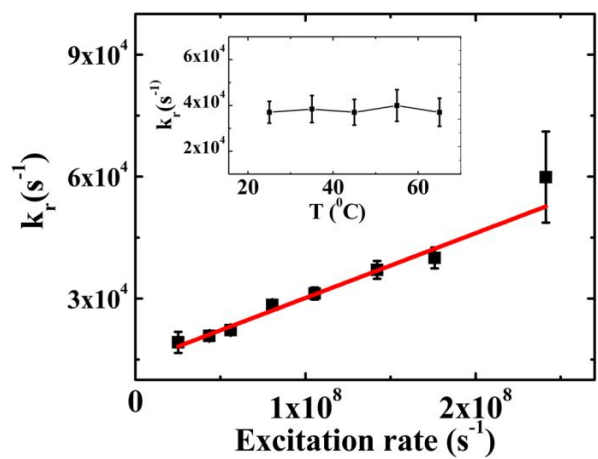

**Figure 3:** Dependence of $k_r$ on excitation rate. The continuous line is a straight line fit with a slope 2.4 X $10^{-4}$. The inset shows $k_r$ verses temperature.

**Figure 4.**

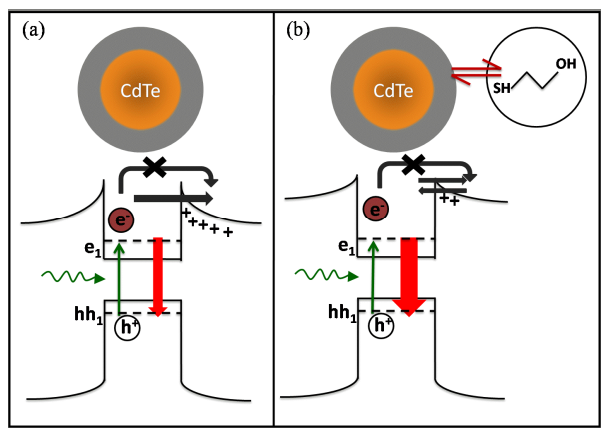

**Figure 4.** A schematic diagram describing the charge transfer process in photo-excited QD. a) The transfer of electron without the BME addition. The electron tunnels through the barrier upon absorbing a photon with probability ≈ $10^{-6}$, The neutralization rates are negligibly small compared to the ionization rates and overall photo-luminescence is reduced. The thick vertical arrow represents radiative recombination. The horizontal arrows represent the ionization. b) After addition of BME the ionization probability per photon increases to ≈ $10^{-4}$.